# MODELING THE COMPUTER MEMORY BASED ON THE FERROMAGNET/SUPERCONDUCTOR MULTILAYERS

S. E. Shafraniuk[2], I. P. Nevirkovets[1], and O. A. Mukhanov[1]
[1]Hypres, Inc., 175 Clearbrook Road, Elmsford, NY 10523, USA and
[2]Tegri LLC, 558 Michigan Ave, Evanston, IL 60208, USA

*(Dated: 2018)*

**A model of superconducting computer memory exploiting the orthogonal spin transfer (OST) in the pseudospin valve (PS) that is controlled by the three-terminal Josephson superconducting-ferromagnetic transistor (SFT) is developed. The building blocks of the memory are hybrid PS and SFT structures. The memory model is formulated in terms of the equation-defined PS and SFT devices integrated into the PS/SFT memory cell (MC) circuit. Logical units "0" and "1" are associated with the two PS states respectively characterized by two different values of resistance. Elementary logical operations comprising the read/write processes occur when a word pulse applied to the SFT's injector coincides with the respective bit pulse acting on MC. Physically, a word pulse triggers SFT to a resistive state, causing the PS switching between the logical "0" and "1" states. Thus, the whole switching dynamics of MC depends on the non-equilibrium and non-stationary properties of PS and SFT. Modeling the single MC as well as the larger MC-based circuits comprising respectively twelve and thirty elements suggest that such the memory cells can undergo ultrafast switching (sub-ns) and low energy consumption per operation (sub-100 fJ). The suggested model allows studying the influence of noises, punch-through effect, crosstalk, parasitic, etc. The obtained results suggest that the hybrid PS/SFT structures are well-suited to superconducting computing circuits as they are built of magnetic and non-magnetic transition metals and therefore have low impedances (1-30 Ω).**

## 1. Introduction

Issues of the local overheating and thermal management in the contemporary digital semiconducting circuits, serving as an element base for nowadays computers, represent the major impediment to the progress in this area[9]. Presently, computing information is processed by transmitting the electric pulses that trigger logic elements in a circuit involving semiconducting diodes and transistors. Such the elements, when subjected to electric signals, release Joule heat. In large circuits, comprising ~$10^6$-$10^8$ elements per chip, the total amount of the released heat is significant, hampering the whole computer's performance[10]. Today, as power supply for a supercomputer typically a small-sized power plant is needed. The overheating issue worsens when the clock frequency and the element density in the circuits increase. One promising solution to this long-standing problem obstructing the computing efficiency and speed is development of the Josephson-based electronics [9,11-14], whose basic physical principles and functionality are quite different from its semiconducting transistor counterparts. Although in the both materials, semiconductors and superconductors, there is an energy gap in the quasiparticle excitation spectrum, the quantum coherence of the superfluid condensate of the latter envisages remarkable properties. The benefit is that the finite voltage state of the Josephson junction (JJ) is not related to energy dissipation, thus no Joule heat is released. The finite voltage across JJ results from the time-dependent change of the condensate's phase difference occurring without the energy dissipation. For such the reason, the energy dissipation does not impact the clock speed of the superconducting electronic circuits as it happens in its semiconducting counterparts. Since the superconducting wires and logic elements release much lower heat compared to their normal and semiconducting counterparts, energy efficiency of the computing process is improved by several orders of magnitude. Furthermore, the switching speed of the Josephson-based logic elements is much higher than the semiconducting devices[9,11-14]. Besides, recent progress in improving the cooling efficiency and thermal management[15] based on the newly discovered two-dimensional materials makes superconducting computing even more promising.

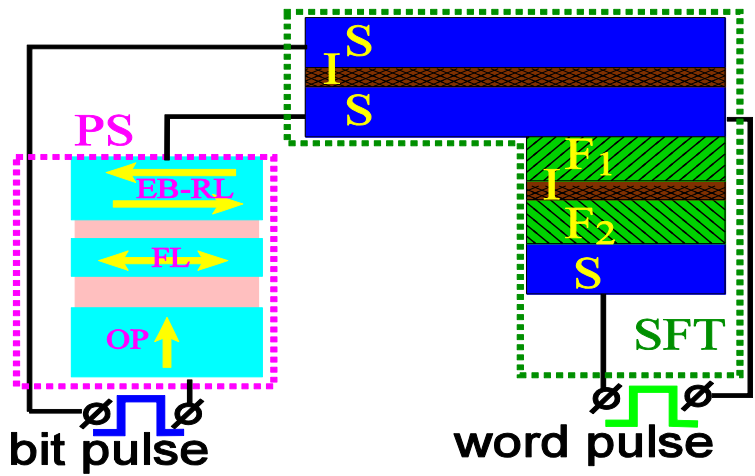

Fig. 1. Computer memory cell (MC) based on the pseudospin valve (PS) controlled by the superconductor-ferromagnetic transistor (SFT). Here EB-RL is the exchange biased synthetic antiferromagnet, FL is the in-plane magnetized soft free layer controlled by the out-of-plane polarizer (OP).

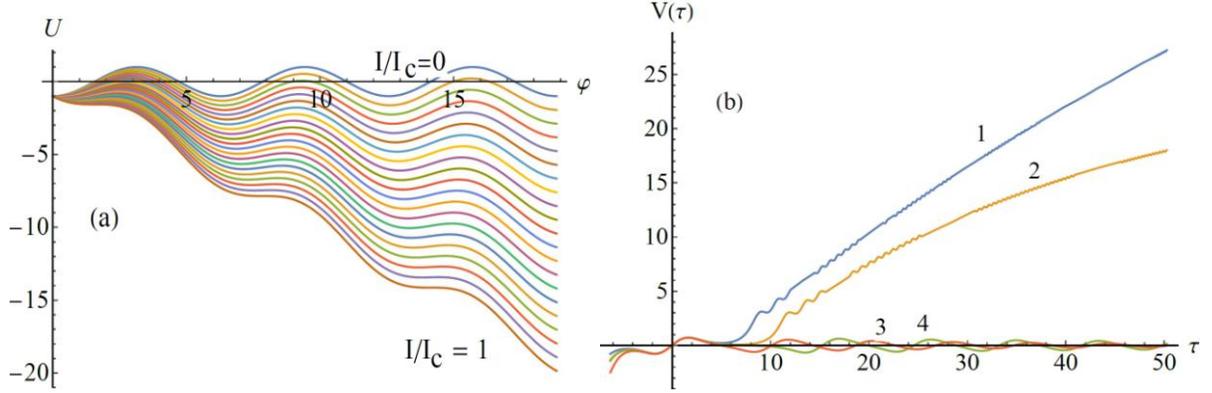

Fig. 2. Non-stationary properties of SFT. (a) The washboard potential $U(\varphi)$ that tilts stronger as the bias current increases from $I/I_c = 0$ up to $I/I_c = 1$. In SFT, one changes tilt either by adjusting $I$ or $I_c$ achieving much better flexibility of control, as compared to a conventional Josephson junction, where the tilt is controlled only by $I$. (b) The time dependence of the JJ voltage $V(\tau)$, where $\tau = \omega_J t$ is the dimensionless time, $\omega_J$ is the Josephson plasma frequency. Curves 1, 2, 3, and 4 are related respectively to $\beta_J = 0.01, 0.03, 0.033$ and $0.065$. We emphasize that in SFT one can readily alter both, $\omega_J$ and $\beta_J$ by merely changing the injector current.

The superconducting and semiconducting electronics[9,11-14] are based on distinct physical principles. In particular, present superconducting circuits are typically designed using the Josephson junctions just with two terminals, and absence the superconducting transistor element imposes restrictions on the overall circuit's performance and also limits their capabilities. Therefore, on the path to creating the fully functional and efficient superconducting computer there remain important technical issues requiring further elaboration. One of the key problems is development the three-terminal Josephson transistor[1-4] that would essentially simplify designing the whole superconducting electronics. In particular, the Josephson transistor, when available, would essentially diversify the schematics and improve the overall performance, since the digital logic based on the three-terminal Josephson transistors[1-4] would benefit from diminishing the fan out effect, noises and errors, hampering the performance of the currently available Josephson digital circuits [9,14,16,17].

Another emerging problem is development the high-density, speedy, energy-efficient non-volatile and non-destructive superconducting computer memory. Presently known solutions either fail to function as non-volatile, non-destructive memory or they lack satisfactory energy efficiency, speed and stability. Existing computer memory is based on the semiconducting transistors consuming considerable energy and switching at limited speed that impairs its overall performance. Furthermore, when coupled with superconducting circuits, semiconductors become a source of noise and errors.

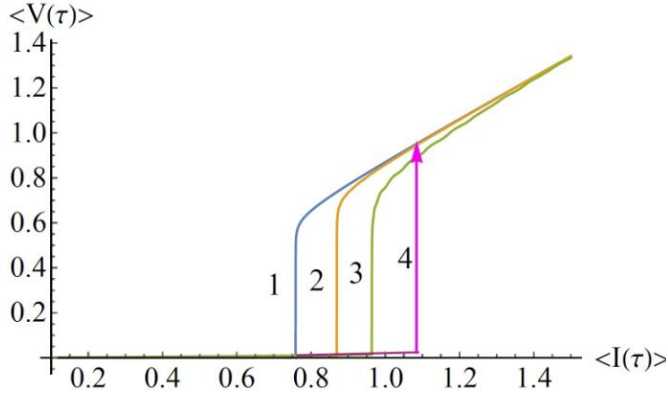

Fig. 3. The time-averaged curves of the SFT acceptor d.c. return current $\langle V(\tau)\rangle$ versus $\langle I(\tau)\rangle$ for $\beta_J = 0.05$ (curve 1), $\beta_J = 0.3$ (curve 2) and $\beta_J = 0.6$ (curve 3). One can see the hysteresis (curve 4) that is larger for smaller $\beta_J$.

In this paper we consider a circuit model of the computer memory (see Fig. 1), whose elements are built using the pseudospin valve (PS) coupled with the three-terminal Josephson transistor (SFT), based on the hybrid superconductor-ferromagnetic multilayer[1-3,18]. The model is elaborated using our previous experimental measurements and theoretical study of the PS[5-8] and SFT structures[1-3,18]. The experimental and theoretical data[1-3,5-8,18] are reformulated here in terms of the equation-defined devices (EDD) comprising the PS/SFT memory cell (MC) circuit. We describe the physics of the invented devices, SFT and pseudospin valve in terms of small phenomenological models, as explained in Secs. 2-4. We will see that such the models (see Secs. 2-4) are more suitable to studying the practical applications in large circuits as the previous microscopic models [18,30,31] are too cumbersome and are good only for single devices. The aims of this work are as follows. (*i*) Introducing the compact models of SFT and pseudospin valve to improve the design capabilities of large superconducting circuits. (*ii*) Tailoring the smaller phenomenological models to matching the complex microscopic theory and the experimental data [18,30,31] that are impractical in the large circuit simulations. (*iii*) Simplifying the phenomenological models as much as possible to increase the speed and quality of large circuit simulations. The pseudospin valve (PS) functionality exploits the phenomenon of orthogonal spin transfer [5-8] (OST), initialized by the sub-nanosecond pulses of the bias electric current altering the magnitude of pseudospin valve resistance $R_{PS}$ between the minimum $R_L$ and maximum $R_H$. Such the PS switching is controlled using the non-equilibrium quasiparticle injection to suppressing the critical Josephson supercurrent in the superconductor-ferromagnetic transistor (SFT). Thus, the STF injector acts as the base electrode of transistor, while the two superconducting S layers of the SIS sub-junction correspond to emitter and collector.

Circuit simulations allow finding the optimal parameters of the hybrid memory cell that ensure the best memory performance. The two PS states respectively characterized by $R_L$ and $R_H$ are associated with the logical units "0" and "1". Reading/writing the logical information is performed by applying a word pulse to the SFT's injector. On the one hand, the non-equilibrium quasiparticle injection suppresses the superconductivity in the acceptor SIS junction, thereby reducing the magnitude $I_{ca}$ of its critical supercurrent. Thus, by applying a word pulse to the SFT's injector, one switches the SIS acceptor into the resistive state. On the other hand, the SIS acceptor is coupled to the PS valve that is subjected to the bit pulses. When the word and bit pulses act simultaneously, the cumulative transport current becomes sufficiently strong to alter the PS valve state, forcing the free layer (FL) to change its polarization in respect to the in-plane polarized reference layer (RL) from parallel (P) to antiparallel (AP), or vise versa. Below we will see that the SIS switching from superconducting to resistive state (or vise versa) can be exploited

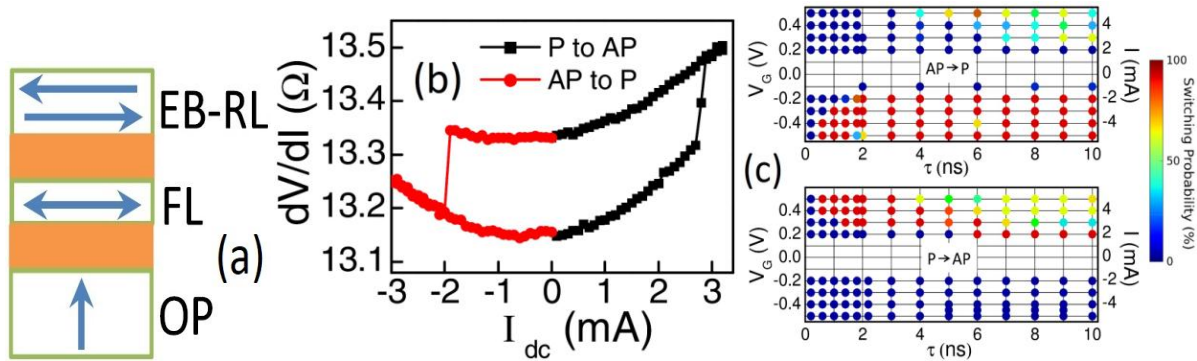

Fig. 4. Experimental data used to building simplified phenomenological models the PS valve [5]. (a) The PS valve, where the out-of-plane polarizer (OP) controls magnetization of the free layer (FL) coupled to the in-plane magnetized reference layer (EB-RL). (b) The OST switching diagram [5-8] modeled by the equation-defined device EDD, according to Eqs. (18)-(21). (c) Switching probability diagram [5] versus voltage $V_G$ (y-axis) and duration $\tau$ (x-axis) mimicked by Eq. (25). The estimated pulsed current ($I$) at the device is shown on the second y-axis. The red color marks the maximal (100%) switching probability whereas the dark blue color means the absence of switching events. Reproduced with permissions from AIP, Ref. [5].

to controlling the PS resistance.

Thus, a necessary condition for the reading/writing process to occur is that the word pulse coincides with the respective bit pulse acting on MC. Physically, a word pulse triggers the Josephson SIS sub-junction part of SFT to a finite voltage (resistive) state, causing the PS switching between the logical "0" and "1" states, manifesting the elementary read/write process. The whole switching dynamics of the PS/SFT memory cell depends on the non-equilibrium and non-stationary properties of the structural components. Other important factors determining the switching dynamics are the geometry and design of the memory circuit that must be elaborated and optimized to reduce parasitics, crosstalk, and noises along with improving the overal thermal management. The goal of this work is studying the non-stationary dynamics of the PS/SFT memory using WRSpice and QUCS simulators complemented by the Wolfram Mathematica and Matlab toolboxes. Below we perform numeric simulations the single PS/SFT memory cell and the larger computer memory circuits comprising twelve and thirty elements.

## 2. Modelling the PS and SFT devices

Design, validating and optimizing the electronic circuits is accomplished using the computer aided design (CAD) and circuit simulations. Therefore, the CAD tools serve to development the semiconducting integrated circuits. However, the CAD tools are not well mature yet for working with the superconducting digital electronics [19]. When new devices are invented, one should apply significant efforts to properly define them and integrate into the electronic circuits. Fabricating the devices and even explaining their physics with complex microscopic models does not guarantee they will properly function in the circuits. This motivates verifying whether or not such the devices can function as elements of the large superconducting circuits. Nonetheless, the circuit simulations cannot run without simple but adequate phenomenological models describing basic physical properties of the SFT and pseudospin valve devices. Since the working principles of the devices are very different from the physics of traditional electronic elements, the answer is far non-trivial. This work aims to show that such the devices, as SFT and pseudospin valve are able to work as elements of the superconductive electronic circuit. In the following Sections we will built the required phenomenological models matching the microscopic theories [18,30,31] and fulfilling requirements of the circuit simulators. The agility and simplicity of such the phenomenological models will allow simulating quite complex superconducting memory circuits, which would be impossible when using the former microscopic models. The compact models of the three-terminal SFT and the two-terminal pseudospin valve thus allow demonstrating that such the devices are suitable for practical applications as they can function as key elements of the large superconducting circuits: SFT governs switching the pseudospin valve posing as a memory cell of the superconducting computer.

A variety of superconducting digital logic devices are based on the RSFQ technology [9,16,17] that is known almost for three decades. But the progress in the RSFQ technology was slowed by shortage of necessary circuit elements. One example was absence of three-terminal superconducting transistor that complicated designing the electric pulse generators, triggers, and switches. These motivated efforts toward development of efficient simulation methods would facilitate successful progress in this field. There were numerous attempts to elaborate the superconducting electronics by adopting simulators, traditionally used for development the semiconducting circuits. Most popular and available CAD tools[19] for superconducting electronics involve the SPICE simulator [20,21] serving as a platform to design and implement models for the Josephson-based superconducting electronic circuits.

Here we implement similar approach to elaborating models of the superconductor-ferromagnetic transistor [1-4] (SFT), of the memory cell (MC) made of the hybrid SFT/PS device depicted in Fig.1 and of the larger superconducting memory circuits. The PS and SFT devices, representing the key parts of MC, behave as highly non-linear elements with essential reactive impedance components, i.e., either as capacitive or inductive. Orthogonal spin transfer pseudospin valve nano-pillar devices [6,22-29] containing an out-of-plane magnetized polarizing layer and in-plane magnetized spin-valve structure show promise as cryogenic magnetic memory elements [30-33]. The PS valves exhibit ultrafast switching (sub-ns) and low energy consumption per operation (sub-100 fJ) owing to large initial spin-transfer torque from the perpendicular polarizer. Furthermore, these devices are well-suited to integration with Josephson junction circuits as they are comprised of magnetic and non-magnetic transition metals and therefore have low impedances (1-30 $\Omega$). However, because PS has only two terminals, it cannot be suitably controlled that limits its functionality as a memory circuit element. In this work, we extend the PS capabilities by integrating it with SFT that result in a fully functional memory cell. First, we formulate and study the models of single PS and SFT elements shown in Figs. A1, A2 (see Appendix). In the next step, we integrate the two models to derive a model of MC. The functional model of MC is used then to subsequent development of larger MC-based memory circuits comprising 12 and 30 elements that are well suited for studying the basic functionality of the PS/SFT computer memory. Schematics of the PS and SFT blocks developed in either graphical or text form in the SPICE[21,34] and QUCS[35-37] simulation environments [38,39], is converted to the system of linear or non-linear differential equations that effectively are the resultant of all the individual components in the circuit. In this way, every circuit component in the simulator is quoted as the compact model of the circuit. The simulations parameters are the coefficients of these set of equations. Mathematically, the equations are deduced from the nodal analysis and the software exploits numerical models of the equation systems to solve them focusing on the specified circuit analysis. Respective compact models are derived for every kind of the component analysis, either transient, DC, AC and temperature, aiming to be represented in a suitable form. Such the form is then used for circuit solutions in SPICE and QUCS[38,39]. Below we describe how to formulate the model for SPICE and QUCS by including the defined non-linear devices into the circuit and solving the system of equations in the circuit setup. We conduct the numerical analyses and find circuit solutions allowing understanding and implementing the device along with the other circuit components under arbitrary conditions that are discussed, without considering particular details of the simulation tools or coding methods.

3. **SFT Josephson transistor**

To accomplish the switching between logical states of the two-terminal PS valve, we complement it with a three-terminal superconducting-ferromagnetic transistor (SFT), whose electrical circuit diagram is shown in Fig. A1 (in Appendix). Namely, the two devices, PS and SFT, are combined into a hybrid electronic circuit comprising the computer memory cell, whose logical states are triggered using SFT introduced and studied in Refs. [1-4]. The SFT device [1-4] is modeled as electronic circuit with non-linear elements in the form of analytical expressions for the electric current $I = I_R + I_C + I_J$ that consists of the active quasiparticle $I_R = V_R/R$ and reactive capacitive $I_C = C(dV_C/dt)$ components complemented by the Josephson supercurrent

$$I_J = I_c \sin\varphi = I_c \sin\left\{\frac{2\pi}{\Phi_0}\int V_J(t)dt\right\} \quad (1)$$

where $\varphi$ and $V_J$ respectively are the superfluid condensate phase difference and the voltage across the junction, $\Phi_0$ is the flux quantum and we have used the Josephson relationship

$$\frac{d\varphi}{dt} = \frac{2\pi}{\Phi_0}V_J(t). \quad (2)$$

It is instructive to represents the Josephson supercurrent in another form as

$$I_J = \frac{\int V_J(t)dt}{L_J(V_J, I_J)}, \quad (3)$$

where we have introduced the effective non-linear inductance

$$L_J = \frac{\Phi_0}{2\pi I_J}\arcsin\left(\frac{I_J}{I_c}\right). \quad (4)$$

In the above formulas, the resistance $R$, the capacitance $C$, and the inductance $L_J$ of the SIS sub-junction (see Fig. 1), which represents the component of SFT are non-linear functions of the electric currents $I_i$ and bias voltages $V_i$ (where $i = R, C, J$), related to respective branches the SFT circuit.

The electric current through the SIS sub-junction of SFT is described by the RCSJ model in terms of the second order differential equations for $\varphi$. In the SIS, the system oscillates in a potential well, giving rise to sinusoidal current description given by equation (1). The phase-dependent Josephson junction energy $U(\varphi)$ takes form of the washboard potential, whose slope increases with the bias current $I$ as shown in Fig. 2, left. Flexibility of control in SFT is accomplished that one changes tilt either by adjusting $I$ or $I_c$, which is impossible in a conventional Josephson junction, where the tilt is controlled only by $I$. Furthermore, in SFT one can readily alter both, $\omega_J$ and $\beta_J$ by merely changing the injector current. When applying enough bias current, the height of the potential well is decreased and the system rolls off to lower potential wells, losing energy due to dissipation. When $I$ is small, the system can get into next potential well owing to the phase diffusion, meaning that the thermal activation is responsible for crossing the energy barrier by the particles [14]. The total electric current through SIS is

$$I = I_J + I_R + I_c \tag{5}$$

that gives the differential equation in the form

$$I_c \sin\varphi + \frac{V}{R_N} + C\frac{dV}{dt} = I. \tag{6}$$

To compute the DC current-voltage characteristic we need to know the time averaged voltage

$$\langle V \rangle = \frac{\hbar}{2e}\left\langle\frac{d\varphi}{dt}\right\rangle = I_c R_N \beta_J \left\langle\frac{d\varphi}{d\tau}\right\rangle, \tag{7}$$

where $I_c$ is the critical current, $\tau = \omega_J t$ is the dimensionless time, $R_N$ is the normal state resistance,

$$\beta_J = \frac{1}{\omega_J R_N C}, \tag{8}$$

and the $\omega_J$ is the Josephson plasma frequency

$$\omega_J = \sqrt{\frac{2eI_c}{\hbar C}}. \tag{9}$$

The McCumber parameter $\beta_c$ is

$$\beta_c = \frac{1}{\beta_J^2} = \frac{2e}{\hbar}I_c R_N^2 C. \tag{10}$$

A simple analytical expression is obtained when the SIS sub-junction capacitance is small. In this limit $\beta_J \gg 1$ and the above Eq. (6) is reduced to

$$\beta_J \frac{d\varphi}{d\tau} + \sin\varphi = \kappa \tag{11}$$

where $\kappa = I/I_c$ and

$$\langle V \rangle = I_c R_N \beta_J \left\langle \frac{d\varphi}{d\tau} \right\rangle = I_c R_N \beta_J \frac{1}{T} \int_0^T \frac{d\varphi}{d\tau} d\tau = 2\pi I_c R_N \frac{\beta_J}{T}, \quad (12)$$

where the period $T$ is defined by

$$T = \frac{2\pi}{\sqrt{\kappa^2 - 1}} \beta_J. \quad (13)$$

Using that

$$T = \int_0^T \frac{d\varphi}{\left(\frac{d\varphi}{d\tau}\right)} = \beta_J \int_0^{2\pi} \frac{d\varphi}{\kappa - \sin\varphi}, \quad (14)$$

we find,

$$\frac{T}{\beta_J} = \frac{2\pi}{\sqrt{\kappa^2 - 1}} \theta(\kappa - 1), \quad (15)$$

that gives

$$\langle V \rangle = 2\pi I_c R_N \frac{\beta_J}{T} = I_c R_N \sqrt{\kappa^2 - 1}. \quad (16)$$

For JJ with an arbitrary capacitance $C$, we solved Eq. (6) numerically. Respective results are represented in Figs. 2, 3. In Fig. 2, left we show the Josephson energy $U(\varphi)$, Fig. 2, right shows the time-dependent voltage $V(\tau)$, in Fig. 3 we present the I-V curve of JJ computed solving Eq. (6) and finding the time average $\langle V(\tau) \rangle$ versus $\langle I(\tau) \rangle$.

In the low-transparent superconducting tunneling junctions, $I_R$ is a non-linear function of $V_R$ with the threshold $V_R = 2\Delta/e$, where $\Delta$ is the superconducting energy gap in the banks of SIS junction [40]. Furthermore, if there are also resistive and capacitance branches of the 3-branch EDD, then

$$I_J = I - I_R - I_s \quad (17)$$

that gives the non-linear inductance of the Josephson junction in the form

$$L_J = \frac{\Phi_0}{2\pi} \frac{1}{I - I_R - I_s} \arcsin\left(\frac{I - I_R - I_s}{I_c}\right), \quad (18)$$

where $I_s$ is the Josephson supercurrent. In the three-terminal SFT, where the SIS Josephson sub-junction is integrated with the non-equilibrium quasiparticle SFIFS injector, as sketched in Fig. 1, the magnitude of the SIS critical current $I_c$ is suppressed owing to the quasiparticle injection [1-4]. The electrical circuit diagram of SFT is shown in Fig. A1 and the respective SFT equation defined device (EDD) subcircuit is depicted in Fig. A2. Thus, when applying Eq. (1) to SFT, we assume that $I_c$ depends on the SFIFS injector bias current $I_i$ as shown in Fig. A2 (right). In our

SFT circuits shown in Figs. A1, A2, the suppression of $I_{ca}$ by the SFIFS injector in Eq. (1) is modeled by an approximate analytical function with a bell shape

$$I_{ca} = \frac{I_{c0}}{1 + \sinh^2\left(\frac{V_{inj}}{V_{wd}}\right)}, \qquad (19)$$

where $V_{inj}$ is the injector bias voltage and $V_{wd}$ is the bell width that are devised from experiments [1-4] or numerical simulations [41,42]. The above Eq. (19) mimics the experimentally obtained curve $\delta I_{ca}/\delta I_i$ shown in Fig. A2, right, as we set $I_i = V_{inj}/R_i$, where $R_i$ is the injector resistance. For the SFT modeling we used typical values $V_{wd} = 2\,\text{mV}$ and $V_{inj} = 0\text{-}20\,\text{mV}$. The circuit parameters in Fig. A1 (in Appendix A1) are selected to maximize the transistor effect when the electric pulses on the source V1 with amplitude $V_1 = 50\,\mu\text{V} - 1\,\text{mV}$ and duration $\tau_p = 0.13$ ps applied to the SFT base cause substantial change of electric potential $V_{\text{Node1}}(t)$ at Node1, as is evident from the respective transient analyzes diagrams shown in Fig. 5. The respective electric current pulses are measured by Pr1. A strong transistor effect is observed for $R_{1,2} = 1\,\Omega$, $R_3 = 13\,\Omega$, electric potential $V_2 = 60\,\text{mV}$ at source V2. The respective sub-circuit parameters in Figs. A1, A2 (see the equation-defined device D1 in Fig. A2, left) are R9=1 MΩ, R6, R7=0.1 Ω, R3=R8=0.8 Ω, C2=0.001 fF.

In the following sections, using the above formulated EDD model of SFT, we elaborate electronic circuits exploiting the Josephson transistor effect. Adding the SFT elements to the superconducting electronic circuits gives considerable benefits. First, the circuitry is greatly simplified; while many unwanted side effects are excluded. Second, a strong advantage of the SFT-based circuits is that they consume much lower energy and can function at much higher clock speed as compared to their semiconducting counterparts. Other alternatives to SFT have been considered earlier in Refs. [45-47]. According to our estimations [18], the SFT design has better capabilities to controlling the other elements in the circuits when working to switch the pseudospin valves in the computer memory as discussed below.

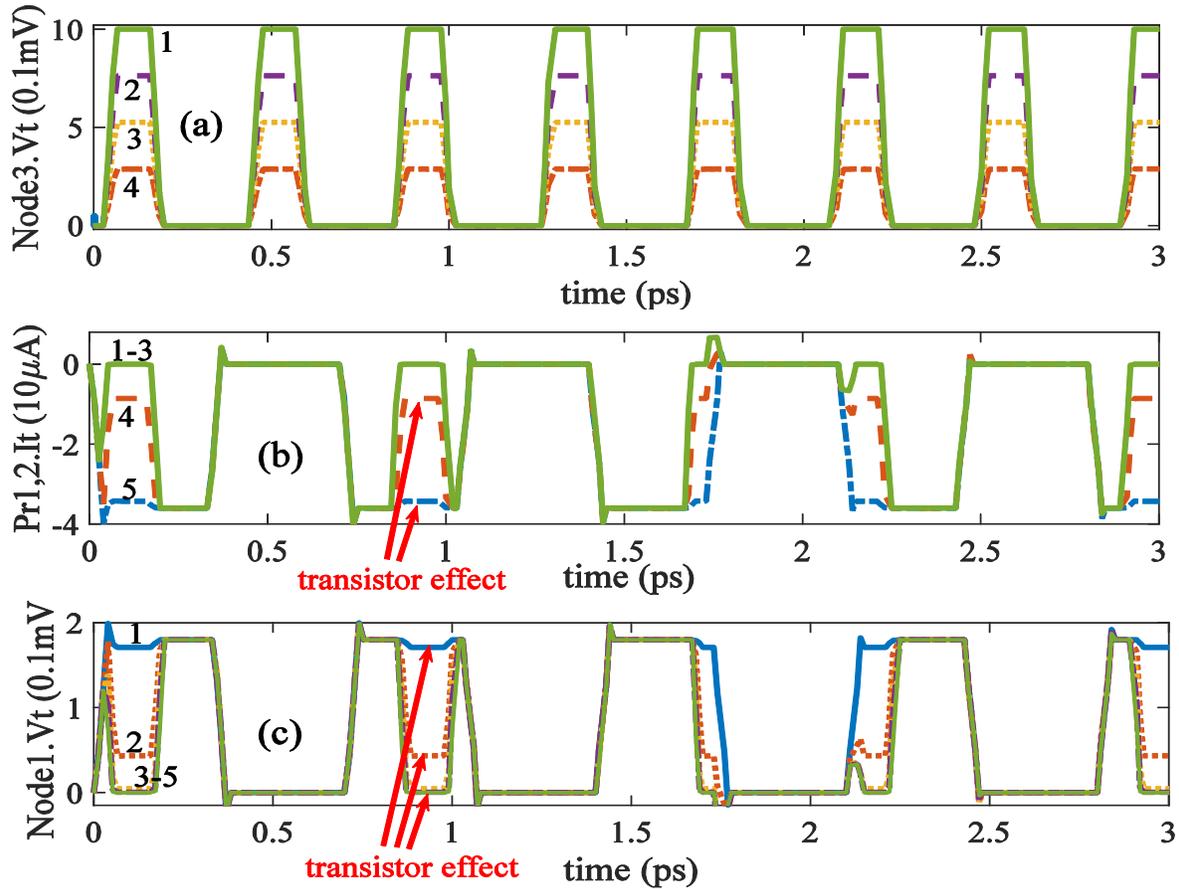

Fig. 5. Transition analyzes diagrams of the test circuit (see Figs. A1, A2) of SFT serving as the Josephson transistor. One notice a strong transistor effect detected by the electric current probes Pr1, Pr2 and voltage $V_1(t)$ at Node1. Curves 1-5 correspond to the voltage pulse amplitudes respectively $U_{in}$ = 5 µV (curve 1), 287 µV (curve 2), 1 mV (curve 3), 5 mV (curve 4), 10 mV (curve 5) at the source V1.

## 4. Hysteresis in the Josephson junctions

The electric current $I$ between two superconductors separated by a weak link forming a Josephson junction flows without dissipation until $I$ reaches a critical current $I_c$, at which a finite bias voltage appears. The I-V curve is described by the RCSJ model (Resistively and Capacitively Shunted Junction) [17]. In conventional SIS Josephson junctions, the hysteresis in the current-voltage characteristic is determined by the junction capacitance $C$ when it is relatively large. Another mechanism of the hysteresis takes place in lateral SNS junctions, where the distance between the two superconducting electrodes results in a small capacitance, much lower than those in typical SIS junction. Then, an observable hysteresis occurs in lateral junctions owing to local heating in the N spacer, when their critical current is large: once the junction has switched to the resistive branch, it does not recover the superconducting state until the bias current is decreased to a significantly smaller retrapping current $I_r$. Such the hysteresis was observed in superconducting constrictions and microbridges (see references, e.g., in Ref. [43]), in superconducting nanowires, normal metals, two-dimensional electron gases (2-DEG), semiconductor nanowires, carbon nanotubes and graphene. Thus, there are two known mechanisms explaining the hysteresis: (i) an excessive Joule heat in the weak link induces there an increase of the local temperature; (ii) the effective capacitance is larger than the geometric capacitance.

Before conducting transient simulations, we briefly discuss the time averaged properties of SFT. In Figs. 2, 3 we represent numeric solutions for the current-biased SFT illustrating its non-stationary properties. From Fig. 2 (left) one can see that the washboard potential $U(\varphi)$ tilts stronger as the bias current increases from $I/I_c = 0$ up to $I/I_c = 1$. The $U(\varphi)$ slope changes either by adjusting $I$ or $I_c$, which provides much better flexibility of control in SFT, as compared to a conventional Josephson junction, where the slope is controlled only by $I$. In Fig. 2 (right) we show the time dependence of the SFT voltage $V(\tau)$, where $\tau = \omega_J t$ is the dimensionless time, $\omega_J$ is the Josephson plasma frequency curves 1, 2, 3, and 4 are related respectively to $\beta_J = 0.01$, 0.03, 0.033 and 0.065. In Fig. 3, we show the time-averaged SFT d.c. return current curves $\langle V(\tau) \rangle$ versus $\langle I(\tau) \rangle$ for $\beta_J = 0.05$ (curve 1), $\beta_J = 0.3$ (curve 2) and $\beta_J = 0.6$ (curve 3). Remarkably, that in SFT one can readily alter both, $\omega_J$ and $\beta_J$ by merely changing the injector current. From this Fig. 3 one can conclude that the I-V curves of such junctions show a hysteresis, which becomes larger for smaller $\beta_J$. In this work we include the I-V curve hysteresis into the circuit model regardless its microscopic mechanism by approximating the quasiparticle branch of the I-V curve by a simple analytical expression, while the Josephson supercurrent is found by solving the RCSJ equations in the course of the circuit simulations. Coefficients of the RCSJ equations that correspond to certain magnitudes of the normal state resistance, capacitance, and critical supercurrent are selected by optimizing the circuit performance in the course of simulations.

## 5. Pseudo-spin valve (PS)

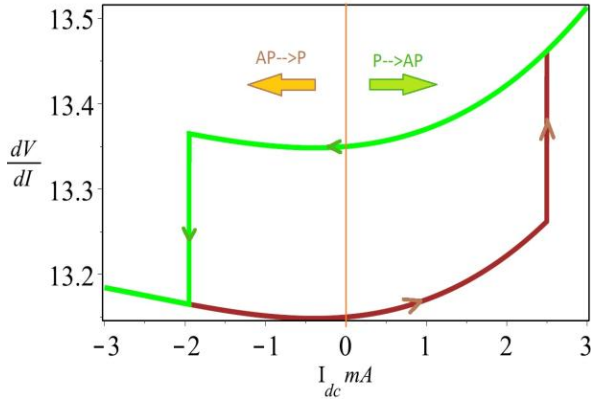

Fig. 6. Model form of the OST switching diagram used in the equation-defined device (EDD). The PS differential resistance $dV/dI_{dc}$ (in Ω) switches step-wise at $I_{dc}$=-2 mA and $I_{dc}$=2.4 mA. The area of the switching loop depends on the pulse duration $t_{pulse}$, whose optimal value is ~5 ns.

The pseudo-spin valve (PS), whose switching is induced by the orthohonal spin transfer (OST) [13-2], controlled by the superconducting-ferromagnetic transistor (SFT), serves as elementary cell of the memory circuit. The pseudospin (PS) valve nano-pillar devices contain an out-of-plane magnetized polarizing layer (OP) and in-plane magnetized spin-valve structure. These devices are compatible with the Josephson junction circuits.

We describe the logical states of superconducting digital circuits in terms of the state variable *St*. For a particular PS element, the initial *state parameter* is defined as

$$St = \begin{cases} 0 \text{ for } \uparrow\uparrow \\ 1 \text{ for } \uparrow\downarrow, \end{cases} \quad (20)$$

where $\uparrow\uparrow$ and $\uparrow\downarrow$ stand respectively for the parallel (P) and anti-parallel (AP) magnetization. As a departure point, we use the experimentally measurered switching diagram of the OST PS valve resistance $R_{PS}(I) = dV(I)/dI$, where $I$ and $V$ respectively are the bias current and voltage, depicted in Fig. 4. The hysteresis in $R_{PS}(I)$ is described in terms of equation defined device (EED). The state variable *St*, related to values of the lower $R_L$ and higher $R_H$ of $R_{PS}(I)$ is set by the magnitude of magnetic field $H$ (or by the bias current). Typically, the PS switching characteristics show a strong hysteresis that must be properly accounted for in the circuit simulations. Specifically, the PS valve state switching forth $0 \rightarrow 1$ and back $1 \rightarrow 0$ occur at different magnitudes of the bias voltage. Therefore, we introduce the two lower $V_{Ll}$, $V_{Lr}$ and two higher $V_{Hl}$, $V_{Hr}$ switching voltages that are established by particular PS valve geometry and material parameters. Simplest analytical form of $R_{PS}$ is written using the step-wise function $\theta(x)$

$$R_{PS}(V) = \theta(V_{Lr}-V)\delta_{S,0}R_L + \theta(V-V_{Lr})\theta(V_{Hr}-V)\delta_{S,0}(R_L + Sl\cdot(V_{Hr}-V)) \\ + \theta(V-V_{Hl})\delta_{S,1}R_H + \theta(V_{Hl}-V)\theta(V-V_{Ll})\delta_{S,1}(R_L + Sl\cdot(V_{Hl}-V)), \qquad (21)$$

where $Sl$ is a finite slope, extracted from the experiment. The respective electric current $I(V)$ through the PS valve then is

$$I(V) = \frac{V}{R_{PS}(V)}. \qquad (22)$$

The above $R_{PS}$ representation (19) illustrates how to include the actual experimental data. The resistance $R_{L(H)}$ in the bias voltage intervals $V_{Ll} < V < V_{Lr}$ and $V_{Hl} < V < V_{Hr}$ is approximated by tailoring the respective piece-wise cubical parabolas describing the bias voltage dependence

$$R_{L(H)}(V) = R_{L(H)}^{(0)} + \alpha_{L(H)}V + \beta_{L(H)}V^2 + \gamma_{L(H)}V^3. \qquad (23)$$

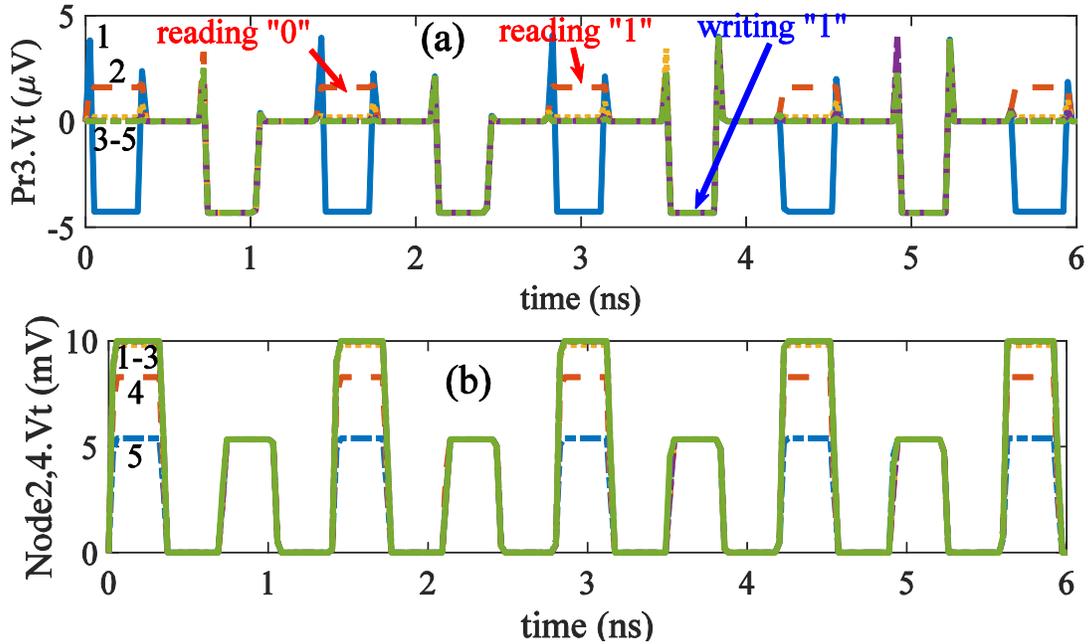

Fig. 7. Transient simulation results for the PS/SFT cell shown in Fig. A3. (a) The diagram for the Pr3.Vt probe shows switching between the positive and negative signals during the read and write "0" → "1" processes. (b) The bit line pulse magnitudes (Node2.Vt and Node4.Vt) are positive. Positive/negative signal of the odd pulses shows reading "0/1", while the positive/negative signal of the even pulses corresponds to writing "0/1", provided the preceding state of PS is "0". To writing "0/1" provided the preceding state of PS is "1", one uses negative bit line pulses (lower graph). Curves 1-5 correspond to the voltage pulse amplitudes respectively $U_{in}$ = 30 µV (curve 1), 273 µV (curve 2), 515 µV (curve 3), 5 mV (curve 4), 10 mV (curve 5) at the source V1.

The coefficients $\alpha_{L(H)}$, $\beta_{L(H)}$ and $\gamma_{L(H)}$ are obtained by fitting the experimental curve shown in Fig. 4, center. The respective model form of $R_{PS}(V)$ is shown in Fig. 6. Along with $R_{PS}$, whose time response to the applied electric bit pulse is step-wise, and the Fourier transform takes the form $R_{PS}(\omega) \propto (1/2)[\delta(\omega) - i/(\pi\omega)]$, one also introduces an effective non-linear inductance

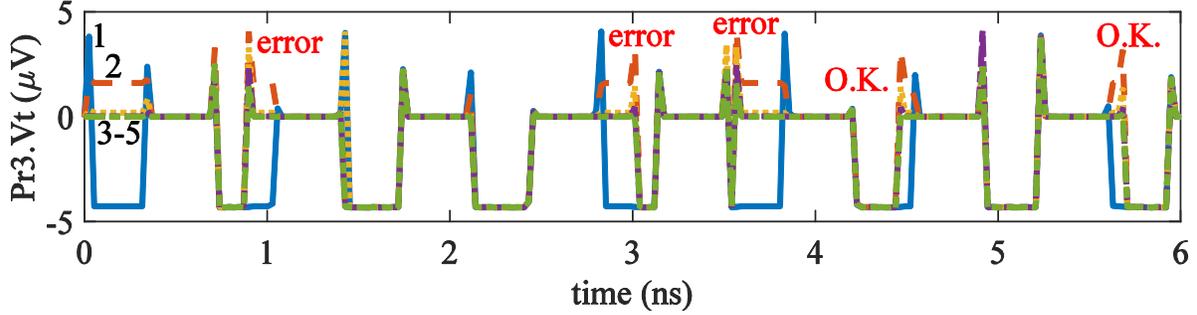

Fig. 8. Example of the transient simulation the OST-SFT memory circuit, showing the readout/writing signal from 2 memory cells Pr1, Pr3 presented in Fig. A4. Curves 1-5 correspond to the voltage pulse amplitudes respectively $U_{in}$ = 15 µV (curve 1), 260 µV (curve 2), 715 µV (curve 3), 5 mV (curve 4), 10 mV (curve 5) that are the same for all the word lines V1,3-13.

$L_{PS}$ by

$$L_{PS}(V,I) I_{PS} = \int V_{PS}(t)dt. \qquad (24)$$

The respective "reactive resistance" $R_{PS}$ for a harmonic signal takes the form $R_{PS}^{(L)} = i\omega L_{PS}$. One estimates that $L_{PS} = 1.6 \times 10^{-3}\,\mu H$ for $\omega = 2\pi f = 2\pi \cdot 0.5$ GHz. Then, the time dependence of the PS valve switching probability is mimicked by introducing a non-linear inductance connected in series with the PS valve. At relatively low $\omega < 10$ GHz, $I_{PS}$ is large enough to ensure the reliable PS valve witching. On the contrary, as $\omega$ increases above ~10 GHz, the electric current $I_{PS}$ through the PS valve is diminished due to the $i\omega L_{PS}$ increase, hence the switching is suppressed. If there are also resistive and capacitance branches of the 3-branch equation-defined device (EDD), then $I_{PS} = I - I_R - I_C$. A simplest equation-defined device (EDD) in this model is introduced using that

$$I_i = \frac{1}{L_{PS}(V_i, I)} \int V_i(t)dt, \qquad (25)$$

where $L_{PS}(V_i, I)$ depends on $i$-th branch voltage $V_i$ and on the total current $I$.

A simplest EDD form of the PS non-linear differential resistance $R_{PS}$ versus the switching current $I_{dc}$ and the pulse duration $\tau_{sw}$ is analytically approximated by

$$R_{OST} = R_P + \frac{1}{2}R_{sw}\tanh\left(\frac{I_{dc}-I_h}{I_{wd}}\right)\left(1+\tanh\left(\frac{\tau-\tau_{sw}}{T_{wd}}\right)\right) \qquad (26)$$

where $I_h$, $R_P$, $R_{sw}$, $I_{wd}$, $\tau_{sw}$, $T_{wd}$ are the PS switching parameters that are devised either using the experimental data [5-8] or the PS valve microscopic modeling results [41,42]. The PS switching event ($0 \rightarrow 1$) happens when the energy supplied by the external power source exceeds a certain threshold value. In particular, such the threshold is evident from the experimental data [5-8]. In the external field $H = 49$ mT, for P $\rightarrow$ AP (i.e., $0 \rightarrow 1$), the switching event occurs when the switching time exceeds 0.6 ns and the bias current exceeds $I_{dc} \geq 2.7$ mA. The backward AP $\rightarrow$ P (i.e., $1 \rightarrow 0$) transition occurs at $I_{dc} \leq -1.9$ mA. The EDD PS valve is represented by respective sub-circuit element defined by Eqs. (20)-(26).

### 6. The PS/SFT memory cell

Further in this work we elaborate the forth $0 \rightarrow 1$ and back $1 \rightarrow 0$ switching processes of the two-terminal PS characterized by two distinct magnitudes $R_L$ and $R_H$ of their resistance[5-8], comprising respectively the two logical states "0" and "1". We have developed model, describing switching dynamics of the PS valve, shown in Fig. 4. The PS switching probability is characterized by a certain critical magnitude of the spin-polarized electron energy $W_c$, transported across the device in the course of the orthogonal spin transfer process. Typical characteristic energies of the two devices, the PS valve and SFT differ by a few orders of magnitude: For the PS valve, the switching energy is $W_{cr}^{PS} \sim 10^{-14} - 10^{-16}$ J while for the SFT device, the respective Josephson energy is $W_{cr}^{SFT} \sim 10^{-19}$ J. The polarization degree of the electron subsystem in the PS valve determines the magnitudes of the minimum $R_L$ and maximum $R_H$ resistance of the OST valve. The model depends on basic physical characteristics of the orthogonal spin transfer and geometry of the PS valve. The OST model circuit is complemented by the SFT circuit shown in Figs. A1, A2, whose switching dynamics is illustrated in Fig. 5 by the respective transient simulation diagrams. To induce the PS switching between the two logical states (20), we exploit suppression of the critical supercurrent $I_c(I_i)$ due to the quasiparticle injection determined by the injector current $I_i$, as illustrated by Fig. A1, right panel, where we show the critical supercurrent $I_{ac}$ and gain $\delta I_c / \delta I_i$ of the SIS acceptor versus the SF$_1$IF$_2$S injector current $I_i$. Physical mechanism of the switching event is that the word pulses applied to SF$_1$IF$_2$S sub-junction, cause injection the non-equilibrium quasiparticles into the SIS acceptor, thereby changing its critical supercurrent $I_{ca}$, as determined by the $I_c(I_i)$ plot in Fig. A1 on inset. This brings SIS into the finite voltage state that also initializes change of the PS resistance $R_{PS}$, provided that simultaneously a bit pulse is applied to PS. Otherwise, when the word and bit pulses do not coincide with each other, the state of PS remains unchanged.

Obtained information was utilized to develop schematics of the PS/SFT memory cell (MC) shown in Fig. 1, whose electronic circuit is shown in Fig. A3. The test circuit, involving the dc simulation, parameter sweep, ac simulation, and transient simulation, allows a detailed study the switching dynamics of the single PS/SFT memory cell. The PS valve is inserted into the Wheatstone bridge, allowing determining the current logical state that is either "0" or "1". Basic functionality of PS/SFT memory cell depends on a variety of factors involving not only the fundamental physical processes taking place in the OST and SFT subsystems, but also comprises design and geometry of the electronic circuit. The transient simulation results are given in the next Fig. 7. One can analyze dynamics of the PS valve switching, reflecting the process of writing and reading the logical information in the PS/SFT memory cell. In the circuit on Fig. A3, the word pulse is applied to the SFT base while the bit pulse biases the emitter-collector line, both governing the read/write logical operations. Such the operations involve a writing process by applying positive pulses for the writing and reading "0" → "1" while the negative pulses are used for the writing and reading "1"→"0". From the test results one can infer that the read/write process in the Fig. A3 computer memory circuit is non-destructive and non-volatile. An important issue that there is a mismatch between the Josephson and PS valve switching energies, which requires using of high magnitudes of $I_{ca}$, of a low shunting capacitance and of low shunting resistance. A straightforward resolution of this issue is accomplished using SNS junctions instead of the SIS junctions. Further increase of the SFT switching power in the SFT geometry is achieved by using stacks of the SNS junctions instead of a single SIS sub-junction.

An efficient performance the single PS/SFT memory cell shown in Fig. A3 is accomplished using the following parameters. The bit and word line impedances both are $Z = 50\,\Omega$, $R1 = 1\,\Omega$, $R2 = 35\,\Omega$, the Wheatstone bridge resistances are $R3, R4, R5 = 5.15\,\Omega$, the lower and upper resistances of the PS valve are $R_L = 5\,\Omega$ and $R_H = 5.3\,\Omega$ respectively. A flawless operation of the PS/SFT memory cell required using the superconducting junctions with a high Josephson energy and low resistance, assuming that the SNS junctions must be used. This implies a high magnitude of the acceptor's critical current $I_{ca} = 0.32\,\text{A}$. Fig. 7 shows the transient simulation results for the PS/SFT cell depicted in Fig. 8→A3. Switching process between the positive and negative signals during the read and write "0" → "1" processes is presented in the upper diagram for the Pr3.Vt probe, provided that the bit line pulse magnitudes (measured at Node2.Vt and Node4.VT) are positive. Positive/negative signal of the odd pulses are related to reading "0/1", while the positive/negative signal of the even pulses corresponds to writing "0/1", provided the preceding state of PS is "0". To exercise the process of writing "0/1" provided the preceding state of PS is "1", one should apply negative bit line pulses (lower graph) in Fig. 7.

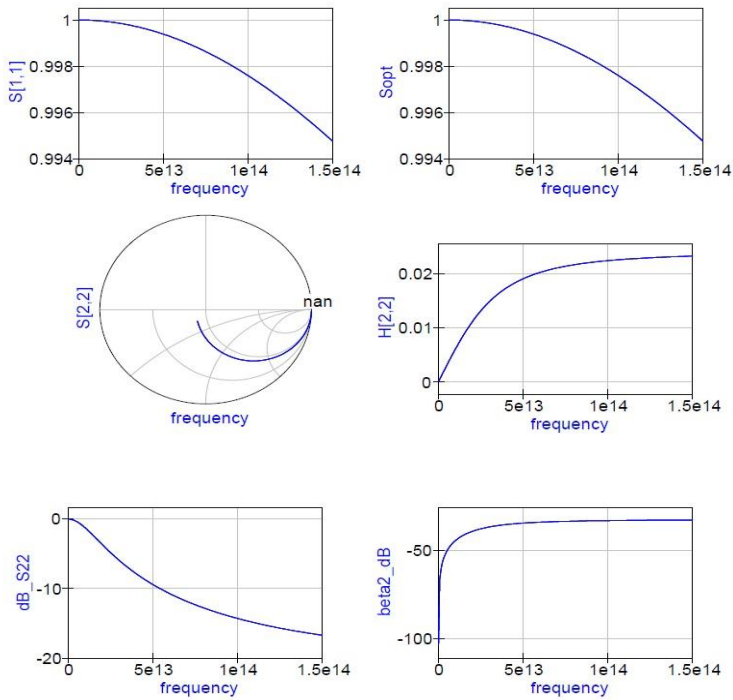

Fig. 9. S-parameter simulation results for the memory cell shown in Fig.8.

To characterize the memory cell further, we compute the S-parameter diagram used to define the signal-wave response of a multiport electrical element at a given frequency. Here, the S-parameter simulation poses as a small-signal AC simulation. It is used to characterize the passive RF component and find the small-signal characteristics of the memory cell at a specific bias. All nonlinear components are linearized and the linear circuit that results is analyzed as a multiport device. Each port is excited in sequence, a linear small-signal simulation is performed, and the response is measured at all ports in the circuit. That response is then converted into S-parameter data, which are in turn sent to the dataset. Respective S-parameter simulation diagrams for circuit Fig. A3 are shown in Fig. 8.

## 7. Computer memory circuit

On the next step, the elaborated models of PS, SFT and MC are integrated into larger and more complex superconducting computer memory circuits. In particular, we modeled the switching dynamics of the computer memory based on the orthogonal spin transfer (OST) involving the pseudospin valves (PS) and the multilayered superconductor-ferromagnetic transistors (SFT) comprising 12 (see the computer memory circuit in Fig. A4) and 30 memory cell elements (not shown). The aim was to understand the basic criterion and establish requirements determining the stability of work and optimal switching parameters of the PS/SFT based computer memory. Simulation results for the memory comprising 12 elements are presented in Fig. 8 as switching diagrams $V_{pr1-9}(t)$. The readout/writing voltage signals $V_{pr1-9}(t)$ shown in Fig. A4 are measured by probes Pr1-Pr9 attached to the individual PS/SFT cells. Schematic of a single PS/SFT memory cell (indicated by the purple ellipses in Fig. A4) was explained earlier in Figs. 7, A4. Furthermore, such the schematic is also evident from Fig. 9 where we show the respective S-parameter simulation circuit and the respective simulation results (Fig. 9). Details of the switching dynamics, in the course of writing and reading the logical information in the MC-based computer memory, critically depend on the circuit parameters that are established based on the present simulations. In the circuit shown in Fig. A4 (see also Fig. A3), the word line electric signal pulses are applied to the SFT base of the cell while the bit line pulses are applied to the emitter-collector line, both governing the read/write operations in the PS/SFT computer memory circuit.

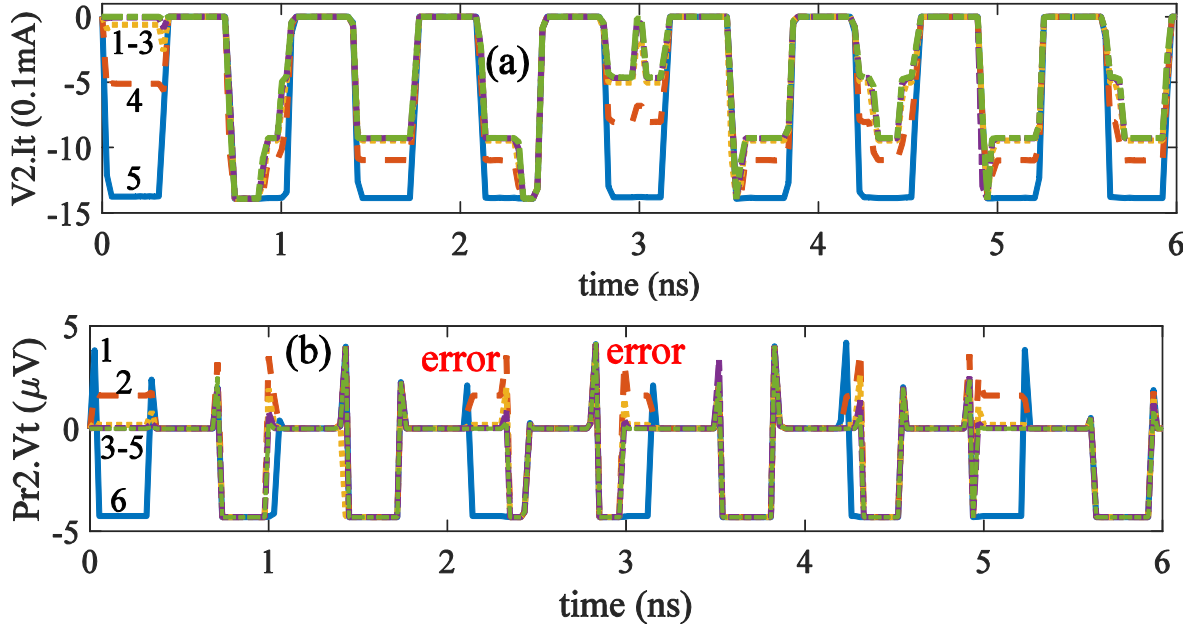

Fig. 10. The transient analyzes diagrams reflecting the time-dependent electric signals. Here V2.It is the electric current at the MC2 emitter while Pr2.Vt is the voltage measured by the Pr2 probe. In order to initialize errors, we set the word pulse period as distinct for different memory cells in the same circuit. Curves 1-5 correspond to the voltage pulse amplitudes respectively $U_{in}$ = 20 μV (curve 1), 252 μV (curve 2), 415 μV (curve 3), 5 mV (curve 4), 10 mV (curve 5) that are the same for all the word lines $V_{1,3,4}$.

The memory circuit operations involve a reading/writing process by applying positive pulses for the writing and reading "0" → "1" and negative pulses for the writing and reading "1"→"0". From the simulation results shown in Fig. 7 it is evident that the read/write process in the memory circuit is non-destructive and non-volatile. The transient analyzes description of the PS/SFT computer memory switching dynamics, although being cumbersome, allows determining basic properties the PS/SFT memory circuit.

### 8. Read/write errors

The origin of errors in the PS/SFT superconductor-ferromagnetic computer memory in the course of reading/writing the logical information has been examined as follows. We utilized the computer-aided design (CAD) of the experimental PS/SFT elements to studying the mechanisms of reading/writing errors in the PS/SFT memory. The errors were intentionally generated by introducing of additional elements and/or by altering parameters the circuit. Particular model parameters depend on the MC geometry and are devised using basic physical principles of the SFT Josephson transistors and the orthogonal spin transfer pseudospin valves. The memory cell (MC) memory model is built by integrating the previously elaborated models of the pseudospin valves and the multilayered superconductor-ferromagnetic transistors (SFT), as was reported earlier. In this work, the sources of errors were studied in the MC-based circuits made of 12 (Fig. A4) and 30 MC elements (not shown). The goal was to find criteria and requirements to preventing the errors in the course of MC-based memory operation. This would increase the stability of work and facilitate setting the optimal switching parameters of the hybrid computer memory.

The quantitative measure of performance and functionality of the memory circuit is the bit-error rate (BER). Basically, there are several factors determining BER. Besides the noise, the errors also arise from the punchthrough effect, crosstalk, parasitic and the Josephson plasma oscillations, especially at higher frequencies. Schematics of the PS/SFT memory, whose individual MC elements are indicated by the purple ellipses in Fig. A4 (for simplicity we show just three MC elements), was given above. In order to mimic the parasitic effects affecting the performance of the memory circuit, we added capacitances C1, C2 and inductances L1-L3, whose values are changed for different simulation runs. In Fig. 10 we present the simulation results the computer memory in the form of transient analyzes switching diagrams $V_{\mathrm{Pr2}}(t)$. The readout/writing voltage signal $V_{\mathrm{Pr2}}(t)$ shown in Fig. 10 is measured by the probe Pr2 attached to the individual MC element as shown in Fig. A4.

Dynamics of the reading/writing errors occurring during the OST/SFT cell switching depends on the circuit parameters that are devised based on the present simulations. The word pulses are applied to the SFT injector serving as the base of the Josephson transistor. For proper MC switching, they must coincide with the bit pulses acting in the emitter-collector lines, thereby the both type of pulses, when acting simultaneously, govern the read/write operations in the OST/SFT computer memory circuit. With intent to initialize errors, we set the word pulse period as being distinct for different memory cells in the same circuit. The reading/writing operations are performed using positive pulses for writing/reading "0" → "1" while the negative electrical pulses are implemented to writing/reading "1"→"0". One can see that the read/write process in the memory circuit is non-destructive and non-volatile, as is confirmed by the simulation results shown in Fig. 10.

The available experimental data in conjunction with the theoretical modeling suggest that the readout/writing processes depend on the value of the spin-polarized electron energy $W_c$, transferred through the OST pseudospin valve, when the bit pulse is applied. The magnitude of BER also depends on relationship between the critical energies for the OST device $W_{cr}^{OST}$ and the respective energies of the SFT device $W_{cr}^{SFT}$. Thus, to reducing BER, the ratio $W_{cr}^{SFT}/W_{cr}^{OST} \sim 10^{-5}-10^{-3}$ must be chosen appropriately. Along with the MC configuration, BER of the PS/SFT computer memory depends on the circuit's impedance that also affects its switching rate. In turn, the magnitude of impedance depends on the geometry and physical parameters of the SFT Josephson transistors and how it couples with the multilayered ferromagnetic PS. In the memory circuit shown in Fig. A4, the bit errors originate from the circuit dynamics and noise. The OST/SFT memory circuits are affected by two different noise sources, the JJ noise of SFT and the noise generated in the normal state elements. The latter noise is the so-called Johnson noise (JN), whose JN current rms value is $i_{rms} = \sqrt{4k_B T v/R}$, where $k_B$ is the

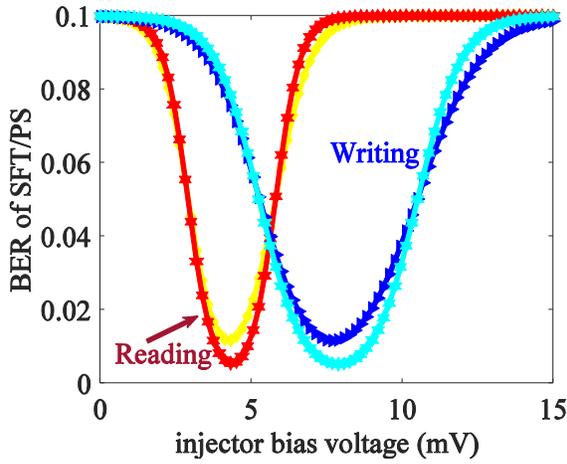

Fig. 11. The bit-error rate (BER) of the memory circuit for reading/writing at $v = 1$ GHz (blue, red) and $v = 1.5$ GHz (cyan, yellow) obtained in the course of the multiple simulation cycles with changing the initial conditions that resulted in a variety of error events.

Boltzmann constant and $v$ is the bandwidth. The JN current consists of the noise current generated inside the MC memory circuit at 4.2 K complemented by the room-temperature noise penetrating from the outside environment along the cables. The noise in a single JJ was examined in Ref. [44], where the authors found that JJ emits a shot noise when the bias voltage $V$ exceeds 1 mV, while the Johnson noise is generated when with $V<1$ mV. We calculated BER of the MC memory circuit in the course of the mixed-signal simulations. The noises are included by adding the respective random sources of current. Furthermore, we considered that the critical current is spread as $I_c = I_c^{(0)} + \delta I_c^{(n)}$, where the randomness $\delta I_c^{(n)}$ depends on the MC index $n$ and is relatively small, $\max \left| \delta I_c^{(n)}/I_c^{(0)} \right| < 0.1$. In the simulations, we also included the randomness of the OSR resistivity difference $\delta R_{PS}^{(n)} = R_{\max}^{(n)} - R_{\min}^{(n)}$, assuming that $\max \left| \delta R_{PS}^{(n)}/R_{\max}^{(0)} \right| < 0.1$. Multiple simulation cycles with changing the initial conditions resulted in a variety of error events. This allowed finding the dependence of BER on the injector bias voltage, corresponding to the magnitude of word pulses on the SFT base as presented in Fig. 11. The obtained results suggest that BER is optimized at certain injector bias voltage $V_i$. This is evident from Fig. 11 where we show BER as function of as a function of the injector bias voltage. One can see that BER is sharply reduced from BER=0.1 down to BER=0.01 for reading at $V_i \sim 4\,\text{mV}$ and for writing at $V_i \sim 8\,\text{mV}$. If the injector bias voltage is insufficient, the omission errors arise, while the insertion errors come up if the injector bias voltage is excessive. Since the Johnson noise prevails, the BER dependence resembles the classical error function, which agrees with the conventional bit-error rate theory.

## 9. Conclusions

The above reported simulation results suggest that the hybrid computer memory is characterized by ultrafast switching (sub-ns) and low energy consumption per operation (sub-100 fJ) due to large initial spin-transfer torque from the perpendicular polarizer. Our model, describing the superconducting computer memory circuits, also allows studying the effect of noises, punch-through effect, crosstalk, parasitic, etc. An advantage of the hybrid PS/SFT devices, serving as elements of the superconducting computing circuits, is that they are built of magnetic and non-magnetic transition metals and therefore have low impedances (1-30 $\Omega$). Therefore, the PS/SFT-based computer memory shows promise in the future's superconducting computers.

Main results of this work are summarized as follows. (i) Creating the small agile phenomenological models adequately capturing the physics of SFT and pseudospin valve making the design of large superconducting circuits possible. In particular we have developed the circuit model of the Josephson transistor. (ii) Proving that using the SFT and pseudospin valve devices adds a lot of remarkable capabilities to the superconducting electronics design. In particular, these proved that the three-terminal Josephson transistor can function as a key element of large circuits and can control other devices in the circuit. This ensures functionality and flawless performance of the computer memory based on the two-terminal pseudospin valve. As an example, we designed the circuit model demonstrating the writing/reading operations in the hybrid PS/SFT computer memory cell (MC) and in larger memory circuits. The model helps optimizing the performance and functionality of the memory circuit characterized by the bit-error rate (BER). (iii) We have examined the criteria and requirements determining the stability of work and switching parameters of the PS/SFT based computer memory. The circuits show remarkably low error rate and good tolerance in large circuit simulations.

## 10. References

[1] I. P. Nevirkovets, O. Chernyashevskyy, G. V. Prokopenko, O. A. Mukhanov and J. B. Ketterson, Control of Supercurrent in Hybrid Superconducting-Ferromagnetic Transistors. Ieee T Appl Supercon 25 (2015).

[2] I. P. Nevirkoves, S. E. Shafraniuk, O. Chernyashevskyy, D. T. Yohannes, O. Mukhanov and J. B. Ketterson, Critical Current Gain in High-j(c) Superconducting-Ferromagnetic Transistors. Ieee T Appl Supercon 26 (2016).

[3] I. P. Nevirkovets, S. E. Shafraniuk, O. Chernyashevskyy, D. T. Yohannes, O. A. Mukhanov, J. B. Ketterson, Investigation of Current Gain in Superconducting-Ferromagnetic Transistors With High-j(c) Acceptor. Ieee T Appl Supercon 27 (2017).

[4] Nevirkovets, I. P., Shafranjuk, S. E., Chernyashevskyy, O., Masilamani, N. & Ketterson, J. B. Current-voltage characteristics of Nb-carbon-Nb junctions. Low Temp Phys. 40, 191-198 (2014).

[5] L. Ye, D. B. Gopman, L. Rehm, D. Backes, G. Wolf, T. Ohki, A. F. Kirichenko, I. V. Vernik, O. A. Mukhanov, and A. D. Kent. Spin-transfer switching of orthogonal spin-valve devices at cryogenic temperatures. J Appl Phys 115 (2014).

[6] Slonczewski, J. C. Current-driven excitation of magnetic multilayers. J Magn Magn Mater 159, L1-L7 (1996).

[7] Berger, L. Emission of spin waves by a magnetic multilayer traversed by a current. Phys Rev B 54, 9353-9358 (1996).

[8] Ye, L., Wolf, G., Pinna, D., Chaves-O'Flynn, G. D. & Kent, A. D. State diagram of an orthogonal spin transfer spin valve device. J Appl Phys 117 (2015).

[9] Mukhanov, O. A. Energy-Efficient Single Flux Quantum Technology. Ieee T Appl Supercon 21, 760-769 (2011).

[10] S. V. Bakurskiy, V. I. Filippov, V. I. Ruzhickiy, N. V. Klenov, I. I. Soloviev, M. Yu. Kupriyanov, and A. A. Golubov, Current-phase relations in SIsFS junctions in the vicinity of 0-pi transition. Phys Rev B 95 (2017).

[11] Kornev, V. K., Soloviev, I. I., Klenov, N. V., Sharafiev, A. V. & Mukhanov, O. A. Linear Bi-SQUID Arrays for Electrically Small Antennas. Ieee T Appl Supercon 21, 713-716 (2011).

[12] Mukhanov, O. A., Kirichenko, A. F., Filippov, T. V. & Sarwana, S. Hybrid Semiconductor-Superconductor Fast-Readout Memory for Digital RF Receivers. Ieee T Appl Supercon 21, 797-800 (2011).

[13] Holmes, D. S., Kadin, A. M. & Johnson, M. W. Superconducting Computing in Large-Scale Hybrid Systems. Computer 48, 34-42 (2015).

[14] Likharev, K. K. Dynamics of Josephson Junctions and Circuits. 1 edn, (CRC Press, 1986).

[15] Shafraniuk, S. Thermoelectricity and Heat Transport in Graphene and Other 2D Nanomaterials. (Elsevier - Health Sciences Division, 2017).

[16] Likharev, K. K. Superconductor digital electronics. Physica C-Superconductivity and Its Applications 482, 6-18 (2012).

[17] Likharev, K. K. & Semenov, V. K. RSFQ Logic/Memory Family: A New Josephson-Junction Technology for Sub-Terahertz-Clock-Frequency Digital Systems. Ieee T Appl Supercon 1, 3-28 (1991).

[18]	Shafranjuk, S., Nevirkovets, I. P., Mukhanov, O. A. & Ketterson, J. B. Control of Superconductivity in a Hybrid Superconducting/Ferromagnetic Multilayer Using Nonequilibrium Tunneling Injection. Phys Rev Appl 6 (2016).

[19]	K. Gaj, Q. Herr, V. Adler, A. Krasniewski, E. G. Friedman, M. S. Feldman, Tools for the computer-aided design of multigigahertz superconducting digital circuits. Ieee T Appl Supercon 9, 18-38 (1999).

[20]	Kalinov, A. V., Voloshin, I. F. & Fisher, L. M. SPICE model of high-temperature superconducting tape: application to resistive fault-current limiter. Supercond Sci Tech 30 (2017).

[21]	Shen, B., Jiang, J. F. & Cai, Q. Y. Simulation of circuits composed of HTSC-single hole transistors by using superconducting SET-SPICE. Physica C 341, 1605-1606 (2000).

[22]	Hernandez, S. & Victora, R. H. Calculation of spin transfer torque in partially polarized spin valves including multiple reflections. Appl Phys Lett 97 (2010).

[23]	Qiu, Y. C., Zhang, Z. Z., Jin, Q. Y. & Liu, Y. W. Dynamic dipolar interaction effect on spin-transfer switching with perpendicular anisotropy. Appl Phys Lett 95 (2009).

[24]	Zhu, X. C. & Kang, S. H. Inherent spin transfer torque driven switching current fluctuations in magnetic element with in-plane magnetization and comparison to perpendicular design. J Appl Phys 106 (2009).

[25]	Lee, K., Chen, W. C., Zhu, X. C., Li, X. & Kang, S. H. Effect of interlayer coupling in CoFeB/Ta/NiFe free layers on the critical switching current of MgO-based magnetic tunnel junctions. J Appl Phys 106 (2009).

[26]	Zhu, X. C. & Kang, S. H. Distinction and correlation between magnetization switchings driven by spin transfer torque and applied magnetic field. J Appl Phys 105 (2009).

[27]	Carpentieri, M., Torres, L. & Martinez, E. Micromagnetic study of spin-transfer driven ferromagnetic resonance: Equivalent circuit. J Appl Phys 106 (2009).

[28]	Carpentieri, M., Torres, L. & Martinez, E. Temperature Dependence of Microwave Nano-Oscillator Linewidths Driven by Spin-Polarized Currents: A Micromagnetic Analysis. Ieee T Magn 45, 3426-3429 (2009).

[29]	Carpentieri, M., Torres, L. & Martinez, E. Effective Damping Contribution From Micromagnetic Modeling in a Spin-Transfer-Driven Ferromagnetic Resonance. Ieee T Nanotechnol 8, 477-481 (2009).

[30]	B. Baek, W. H. Rippard, S. P. Benz, S. E. Russek, & P. D. Dresselhaus, Hybrid superconducting-magnetic memory device using competing order parameters. Nat Commun 5 (2014).

[31]     B. Baek, W. H. Rippard, M. R. Pufall, S. P. Benz, S. E. Russek, H. Rogalla, P. D. Dresselhaus, Spin-Transfer Torque Switching in Nanopillar Superconducting-Magnetic Hybrid Josephson Junctions. Phys Rev Appl 3 (2015).

[32]     C. Bell, G. Burnell, C. W. Leung, E. J. Tarte, D.-J. Kang, and M. G. Blamire, Controllable Josephson current through a pseudospin-valve structure. Appl Phys Lett 84, 1153-1155 (2004).

[33]     M. A. El Qader, R. K. Singh, S. N. Galvin, L. Yu, J. M. Rowell, and N. Newman, Switching at small magnetic fields in Josephson junctions fabricated with ferromagnetic barrier layers, Appl. Phys. Lett. **104**, 022602 (2014).

[34]     Jain, A., Ghosh, A., Singh, N. B. & Sarkar, S. K. A new SPICE macro model of single electron transistor for efficient simulation of single-electronics circuits. Analog Integr Circ S 82, 653-662 (2015).

[35]     W. Grabinski, M. Brinson, P. Nenzi, F. Lannutti, N. Makris, A. Antonopoulos, M. Bucher, Open-source circuit simulation tools for RF compact semiconductor device modelling. Int J Numer Model El 27, 761-779 (2014).

[36]     Brinson, M. E. & Kuznetsov, V. A new approach to compact semiconductor device modelling with Qucs Verilog-A analogue module synthesis. Int J Numer Model El 29, 1070-1088 (2016).

[37]     Brinson, M. E. & Nabijou, H. Adaptive subcircuits and compact Verilog-A macromodels as integrated design and analysis blocks in Qucs circuit simulation. Int J Electron 98, 631-645 (2011).

[38]     Brinson, M. E. & Jahn, S. Qucs: A GPL software package for circuit simulation, compact device modelling and circuit macromodelling from DC to RF and beyond. Int J Numer Model El 22, 297-319 (2009).

[39]     Jahn, S. & Brinson, M. E. Interactive compact device modelling using Qucs equation-defined devices. Int J Numer Model El 21, 335-349 (2008).

[40]     Giaever, I. Energy Gap in Superconductors Measured by Electron Tunneling. Phys Rev Lett 5, 147-148 (1960).

[41]     Pinna, D., Stein, D. L. & Kent, A. D. Spin-torque oscillators with thermal noise: A constant energy orbit approach. Phys Rev B 90 (2014).

[42]     Pinna, D., Kent, A. D. & Stein, D. L. Thermally assisted spin-transfer torque dynamics in energy space. Phys Rev B 88 (2013).

[43]     Courtois, H., Meschke, M., Peltonen, J. T. & Pekola, J. P. Origin of hysteresis in a proximity Josephson junction. Phys Rev Lett 101 (2008).

[44]     Rogovin, D. & Scalapino, D. J. Fluctuation Phenomena in Tunnel-Junctions. Ann Phys-New York 86, 1-90 (1974).

[45] N. Ruppelt, H. Sickinger, R. Menditto, E. Goldobin, D. Koelle, R. Kleiner, O. Vavra, and H. Kohlsted, Observation of 0–p transition in SIsFS Josephson junctions, Applied Physics Letters **106**, 022602 (2015).

[46] A. Moor, A. F. Volkov, and K. B. Efetov, Inhomogeneous state in nonequilibrium superconductor/normal-metal tunnel structures: A Larkin-Ovchinnikov-Fulde-Ferrell-like phase for nonmagnetic systems, Phys. Rev. B **80**, 054516 (2009).

[47] J. J. A. Baselmans, A. F. Morpurgo, B. J. van Wees & T. M. Klapwijk, Reversing the direction of the supercurrent in a controllable Josephson junction, Nature **397**, 43–45 (1999).

## 11. Appendix

### *A1. The test circuit of SFT*

Fig. A1 shows the geometry (left panel) and the electrical circuit diagram (right panel) of the three-terminal superconducting-ferromagnetic transistor (SFT). Mathematical details of the model are given in Sec. 3 of the main text.

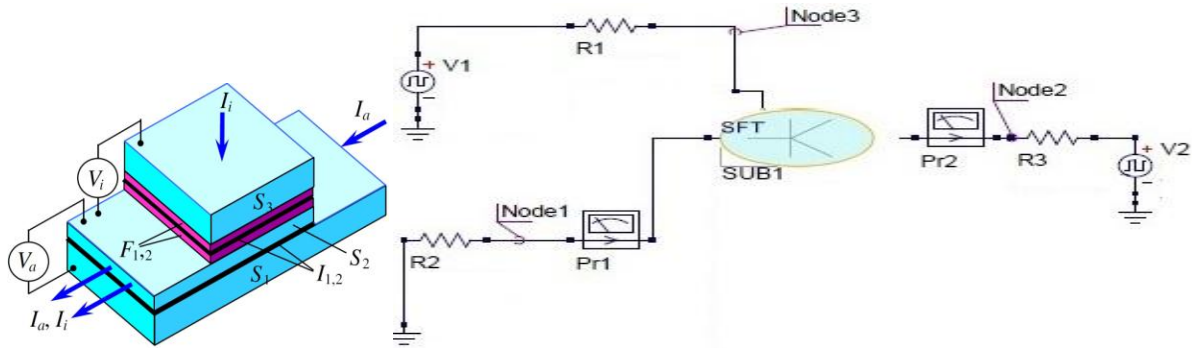

Fig. 5→A1. Test circuit of SFT serving as the Josephson transistor [1-4], whose geometry is shown on the left (reproduced with permissions from APS, Ref. 18). We perform dc simulations, Harmonic balance simulations, ac simulations, and transient simulations for parameters given in text.

### *A2. Respective subcircuit to the SFT equation defined device (EDD)*

The respective SFT equation defined device (EDD) subcircuit is depicted in Fig. A2. In the model by applying the simplified Eq. (1) to SFT, we assume that $I_c$ depends on the SFIFS injector bias current $I_i$ as shown in Fig. A2, right. Detailed circuit parameters are given in main text at the end of Sec.3.

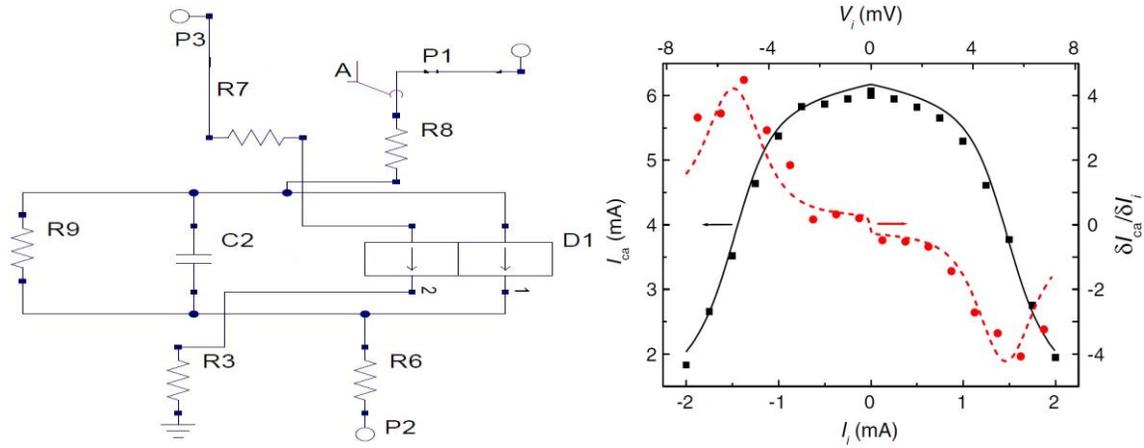

Fig. A2. *Left*: Subcircuit of the SFT Equation Defined Device (EDD) D1 where R9 and C2 are shunting resistance and capacitance respectively. The SFT switching dynamics critically depends on the shunting resistance R9 and capacitance C2. *Right:* Experimental data (points) and theory (solid and dashed curves) used to formulate the EDD, see Eqs. (1), (25). Critical supercurrent $I_{ca}$ (black) and gain $\delta I_c / \delta I_i$ (red) of the SIS acceptor[18] versus the $SF_1IF_2S$ injector current $I_i$. Right panel is reproduced with permissions from APS, Ref. 18.

### A3. Electronic circuit of the PS/SFT memory cell

On the next step, we develop schematics of the PS/SFT memory cell (MC) shown in Fig. A3, whose electronic circuit is shown in Fig. A3. Here SFT serves to accomplish switching between the logical states of the two-terminal PS valve. The two devices, PS and SFT, are combined into a hybrid electronic circuit comprising the computer memory cell, whose logical states are triggered using SFT introduced and studied in Refs. [1-4].

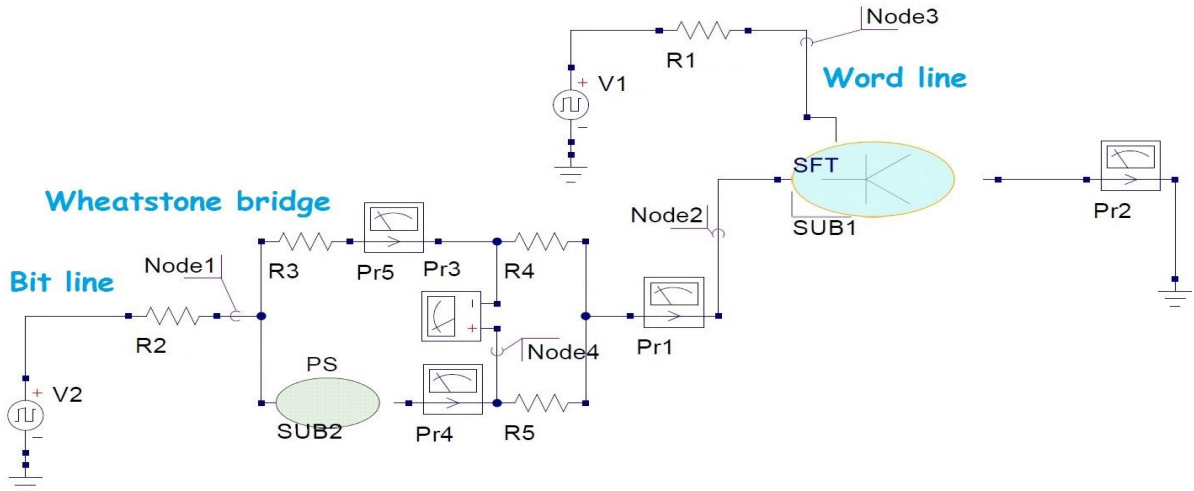

Fig. A3. The PS/SFT computer memory cell devised to studying the switching dynamics. Relevant parameters are indicated near the respective circuit element. The PS valve is inserted into the Wheatstone bridge, allowing determining its current logical state.

Detailed circuit parameters during modeling the single PS/SFT memory cell shown in Fig. A3 are given at the end of Sec. 6 in main text.

### A4. The PS/SFT computer memory circuit comprising 12 cells.

Functional performance of larger computer memory circuit shown in Fig. A4 based on the SFT-controlled pseudospin valves is studied during simulations described in Sec. 7 of main text, where one can find also the details of the circuit design. The individual memory circuit (MC) elements are indicated by the purple ellipses.

Using the developed approach, the elaborated models of PS, SFT and MC are integrated into larger and more complex circuits comprising SFT and PS elements.

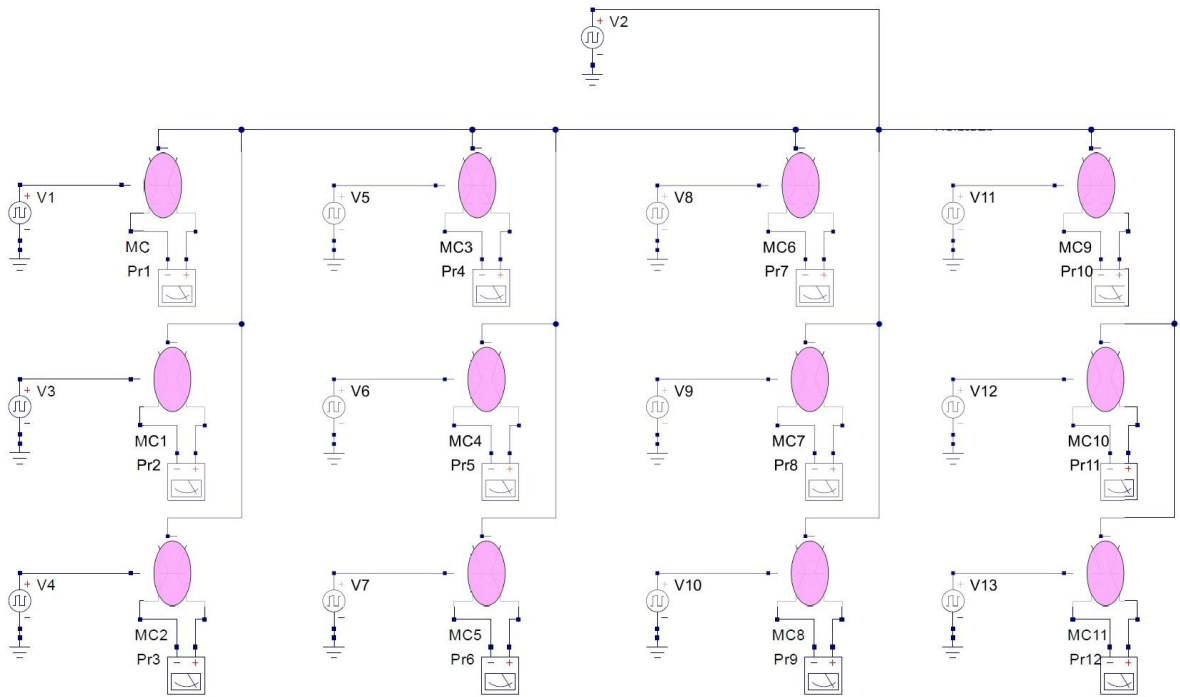

Fig. A4. The PS/SFT computer memory circuit comprising 12 cells. The reading/writing operations are conducted using the multilayered superconductor-ferromagnetic transistors (SFT). To simulate errors, the bit line is common for all the 12 cells while the word lines are individual for each of the cells.

### A5. PS/SFT memory circuit comprising 3 memory cell elements

An important insight into the stability and tolerance of computer memory circuits is devised by mimicking the parasitic effects affecting the performance of the memory circuit. An example of such the test circuit comprising the PS/SFT memory is given in Fig. A5, where for simplicity we show just three MC elements. Detailed circuit parameters and simulation details are given in main text.

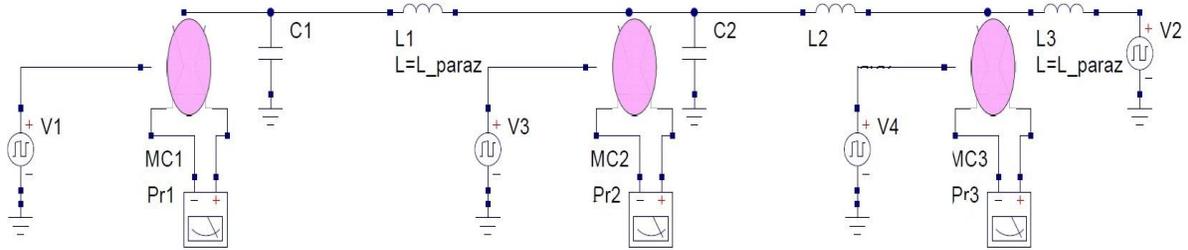

Fig. A5. The PS/SFT computer memory circuit comprising 3 memory cell elements (MC1-MC3) based on the orthogonal spin transfer (OST) controlled by the tree-terminal Josephson SFT transistor. The structure of memory cells and respective sub-circuit schematics were explained in previous reports. The capacitances C1, C2 and inductances L1-L3, whose values are changed for different simulation runs, mimic the parasitic effects. Reading/writing the logical information is performed by applying the word pulses to the SFT base and bit pulse to the emitter-collector.